\documentclass[aps,prl,superscriptaddress,reprint]{revtex4-1}
\usepackage{amsmath} \usepackage{amsfonts}
\usepackage{amssymb} 
\usepackage{array}
\usepackage{graphicx} 
\usepackage{verbatim} 
\usepackage{booktabs}
\usepackage{tabularx}
\usepackage{siunitx} %align by decimal in table

\begin{document}

\title{Thermal Activation Signatures \\ of the Anderson Insulator and the Wigner Solid forming near $\nu=1$ }

\author{S.A. Myers}
\thanks{These authors contributed equally to this work.}
\author{Haoyun Huang}
\thanks{These authors contributed equally to this work.}
\affiliation{Department of Physics and Astronomy, Purdue University, West Lafayette, IN 47907, USA}
\author{Waseem Hussain}
\affiliation{Department of Physics and Astronomy, Purdue University, West Lafayette, IN 47907, USA}
\author{L.N. Pfeiffer}
\affiliation{Department of Electrical Engineering, Princeton University, Princeton, NJ 08544, USA}
\author{K.W. West}
\affiliation{Department of Electrical Engineering, Princeton University, Princeton, NJ 08544, USA}
\author{G.A. Cs\'athy$^\dag$}
\affiliation{Department of Physics and Astronomy, Purdue University, West Lafayette, IN 47907, USA}

\thanks{$\dag$ Corresponding author G.A.C, gcsathy@purdue.edu}

\date{\today}
             
\begin{abstract}

When interactions overcome disorder, integer quantum Hall plateaus support topological phases with different bulk insulators. In the center of the $\nu=1$ plateau the bulk is an Anderson-type insulator, while in the flanks of the plateau the bulk is the integer quantum Hall Wigner solid. We find that the activation energy along the $\nu=1$ plateau exhibits a very dramatic non-monotonic dependence on the magnetic field, a dependence that is strongly correlated with the stability regions of the two phases. 
%An enhanced activation energy in the Wigner solid region is evidence for novel excitations. 
Furthermore, the activation energy has an unexpected minimum at the boundary between the Anderson insulator and the Wigner solid. Our findings constrain the theory of the integer quantum Hall Wigner solid,
determine its thermodynamic properties, and reveal novel behavior at the boundary between the Anderson insulator and the Wigner solid.
\end{abstract}

\maketitle

The effect of interactions in topological matter is currently under intense investigation. Such effects often reveal themselves in low disorder systems, such as the two-dimensional electron gas (2DEG). Integer quantum Hall states (IQHSs) \cite{klitz}, some of the most well-known topological phases, may develop even in the absence of any interaction effects. In contrast, fractional quantum Hall states \cite{tsui} are driven by strong electron-electron interactions, conditions under which the 2DEG is best understood through the formation of emergent particles called composite fermions \cite{jain}. 
Yet members of another wide class of interaction-induced phases retain electron-like particles. Indeed, interactions between electrons in the integer quantum Hall regime can reorganize the insulating bulk into broken symmetry phases, commonly referred to as the electronic bubble and stripe phases \cite{fogler,moessner,lilly,du,cooper,eisen02,zudov,kevin,chen,kivel}. 
The integer quantum Hall Wigner solid (IQHWS), also called the one-electron bubble phase \cite{fogler,moessner}, is one of these broken symmetry phases. 

Besides electronic interactions, the disorder also bears an imprint on how the bulk is ordered.
Depending on the strength of the disorder, the bulk of an IQHS may support distinct insulators. In the limit of strong disorder, the bulk of an integer quantum Hall state has randomly localized quasiparticles \cite{prange,ando}.
Henceforth we will refer to this phase as the Anderson insulator (AI). The AI is characterized solely by topological invariants, or Chern numbers, such as the number of quantized edge states.
In the other extreme of zero disorder, for the bulk of the IQHS theory predicts charge ordered broken symmetry phases, such as the IQHWS \cite{fogler,moessner,kivel}. 
In order to characterize the IQHWS, one needs to identify its Chern number but also order parameters associated with the broken symmetry bulk. IQHWSs were discovered by detecting their pinning modes \cite{chen} and were also revealed by small signal magnetotransport \cite{liu-1,liu-2,sean}, Knight shift \cite{muraki}, chemical potential anomalies \cite{smet}, surface acoustic wave attenuation \cite{suslov}, tunneling \cite{ashoori}, and most recently by high frequency impedance measurements \cite{liu-cap}.
The discovery of the IQHWS in high quality graphene \cite{young} demonstrates  the host-independence of this topological phase and highlights its relevance in a wider class of interacting topological matter. Furthermore, the IQHWS is closely related to the Wigner solid in the extreme quantum limit
\cite{willett-1,gold-1,jiang-1} and other types of Wigner solids that have recently enjoyed a resurgence of interest \cite{reichhardt,denis,hfws-3,hfws-3a,hfws-4,hfws-5,vidhi,ws-6,ws-7,ws-8,joe}. However, in contrast to the IQHWS,
the Wigner solid in the extreme quantum limit does not have edge states associated with it and it does not exhibit quantization. 

All integer quantum Hall plateaus are governed by localization. However, the nature of the localized quasiparticles depends on the interplay of the Zeeman and exchange energies. In the high Landau levels of the GaAs/AlGaAs system, quasiparticles retain a single particle, electron-like nature. In contrast, quasiparticles on the $\nu=1$ plateau carry a non-trivial spin texture and, thus, they acquire a many-body nature.
A quasiparticle with a spin texture is called a skyrmion \cite{sondhi,fertig} and the GaAs/AlGaAs system is especially favorable to skyrmion formation near $\nu=1$ \cite{NMR-Skyrmion,NMR-Skyrmion2,spectroscopy-crystal,poloriazation,NMR-crystal-6,jim,maude}. Therefore for the $\nu=1$ plateau the AI could also be thought of as a phase of randomly localized skyrmions or a skyrmion glass and the IQHWS as a skyrmion crystal \cite{brey-crystal-theory1,green1996,cote-crystal-theory2,Rao1997,  heat-capacitance-crystal,heat-capacitance-crystal2,  resistivitydetectedNMR-crystal,hashi,NMRrelaxation-crystal,  raman-scattering-crystal,  zhu_Skyrmion-crystal}. 
 
Even though there is a considerable literature on both the AI and the IQHWS, much remains unknown about them in the moderate disorder regime, in which both phases develop. Theory either considers the behavior at no disorder \cite{fogler,moessner} or, when disorder was included, the AI so far was not explicitly considered \cite{osw1,osw2,osw3,osw4}. The existence of a boundary between the two phases was only recently pointed out \cite{sean}. The boundary and the transition between a random insulator, such as the AI, and a charge-ordered phase, such as the IQHWS, remain of interest not only in a large variety of condensed matter systems, but also in atomic condensates. 

Here we present a study of thermal excitations of both the AI and the IQHWS forming within the $\nu=1$ integer quantum Hall state. For both the AI and the IQHWS we found thermally activated transport. The activation energy exhibits a very dramatic non-monotonic dependence on the magnetic field. The large variations in the activation energy over the $\nu=1$ integer plateau are in stark contrast to the modest changes in transport. Furthermore, features of the non-monotonic activation energy strongly correlate with the stability regions of the two phases: the activation energy exhibits local maxima in the center of the AI and IQHWS regions and it displays a conspicuous local minimum at the boundary between the two phases. We associate the enhanced activation energy in the IQHWS with the generation of excitations not present in the AI. While the observed phenomena are mainly driven by localization of the two different ground states, the energy scales inferred bear an imprint of exchange correlations and of skyrmion physics.
Our findings constrain the theory of the IQHWS, determine its thermodynamic properties, and reveal new properties of the boundary between the AI and the IQHWS.

We performed low frequency magnetoresistance measurements on exceptional GaAs/AlGaAs samples grown using the latest advances in molecular beam epitaxy techniques \cite{chung}. One of our samples, which we refer to as Sample 1,
has a 2DEG confined to a $75$~nm wide quantum well with an electron density $n=4.2\times 10^{10} \text{ cm}^{-2}$ and mobility $\mu=17\times10^{6}$~cm$^2$/Vs. The sample size is $4\times4\text{ mm}^2$ and it has a total of eight indium ohmic contacts placed at the corners and midpoints of each edge. In addition, we also report thermal activation results in Sample 2 with $n=7.5\times 10^{10} \text{ cm}^{-2}$ and mobility $\mu=24\times10^{6}$~cm$^2$/Vs. Transport and further information on this sample can be found in Ref.\cite{sean}.
Most data were measured in a dilution refrigerator, in which electron thermalization was aided through the use of a He-3 immersion cell \cite{setup}. The largest energy gaps near $\nu=1$ were measured in a PPMS.

\begin{figure}[t]
  \centering
  \includegraphics[width=\columnwidth]{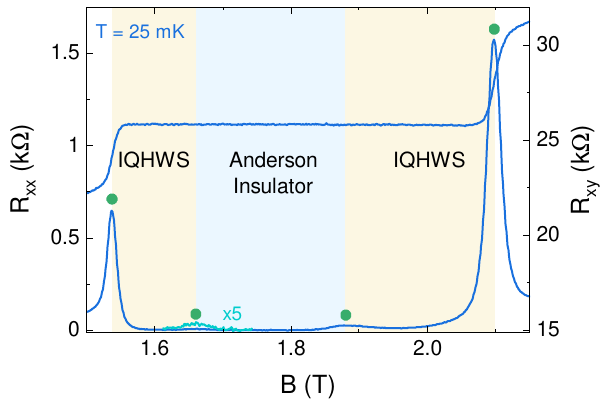}
  \caption{Magnetoresistance $R_{xx}$ and Hall resistance $R_{xy}$ collected from Sample 1 at the temperature of $T=25$~mK. Solid circles are positioned at the local maxima of $R_{xx}$. Near $B=1.66$~T, $R_{xx}$ was multiplied by a factor of five in order to highlight the local maximum developed. Blue and yellow shades mark the stability range of the AI and the IQHWS, respectively.
\label{fig:1} }
\end{figure}

The longitudinal resistance $R_{xx}$ in Sample 1 measured at $T=25$~mK in the vicinity of the $\nu=1$ integer quantum Hall plateau exhibits four local maxima at magnetic fields $B = 1.54, 1.66, 1.88$ and $2.10 $~T. These local maxima of $R_{xx}$ are marked by solid circles in Fig.\ref{fig:1}.
Similar transport features were reported in Refs.\cite{liu-1,sean}.
As seen in Fig.\ref{fig:1}, the four local maxima in $R_{xx}$ split the $\nu=1$ integer quantum Hall plateau in three regions. All three regions have a vanishingly small $R_{xx}$
and $R_{xy}$ quantized to $h/e^2$, hence along the whole plateau the Chern number is $1$. The middle section in the range $1.66$~T $ < B < 1.88$~T straddles $\nu=1$ and it is associated with random localization of the quasiparticles \cite{sean}, or the AI. This region is marked by a blue background in Fig.\ref{fig:1}. In contrast, the $1.54$~T $ <  B < 1.66$~T and $1.88$~T $ <  B < 2.10$~T regions of the $\nu=1$ plateau cannot be due to random localization. These two regions are marked by yellow backgrounds in Fig.\ref{fig:1} and were associated with IQHWSs \cite{liu-1,sean}.
The IQHWS is thought to be pinned by residual impurities \cite{chen,fukuyama,boris-1992,fertig-0}, hence its insulating behavior. The two IQHWSs are linked by particle-hole symmetry \cite{sean}, a pervasive property for a larger family of Wigner solids \cite{vidhi}. 
Transport features of the IQHWS shown in Fig.1 are referred to as the reentrance of the integer quantum Hall effect.
Reentrance was first reported in high Landau levels \cite{lilly,du,cooper,eisen02,zudov,kevin}, where the bulk of the sample was associated with electronic bubble phases \cite{fogler,moessner}. 

IQHWSs were observed near $\nu=1$ in gated two-dimensional electron gases confined to GaAs/AlGaAs quantum wells \cite{liu-1}. These experiments found that the IQHWS develops only above a width-dependent critical density \cite{liu-1}. Ref.\cite{liu-2} found that at the width of $65$~nm, the IQHWS in that experiment developed for densities that exceed $1.4 \times 10^{11}$cm$^{-2}$. In comparison to this finding, in our
sample with a quantum well of $75$~nm the IQHWS develops at a drastically reduced density, $n =4.2 \times 10^{10}$cm$^{-2}$. 
%This density is a factor of $3$ less than the critical density reported in Ref.\cite{liu-2}. 
Such a large density reduction for stabilizing the IQHWS, of about a factor of $3$, most likely resulted from a much higher sample mobility, $17 \times 10^6$cm$^2$/Vs in our sample versus $5\times 10^6$cm$^2$/Vs in Ref.\cite{liu-2}, 
indicating that the formation of the IQHWS is strongly mobility-dependent. 

As pointed out in the introduction, one of the most interesting regions along the $\nu=1$ integer quantum Hall plateau is the boundary between the AI and the IQHWS. At $T=25$~mK, this boundary is determined from the local maxima of $R_{xx}$
near $B = 1.66$ and $1.88$~T \cite{sean}. The filling factors associated with these two magnetic fields, i.e associated with the AI-IQHWS boundary are $\nu_{c,+}=1.06$ and $\nu_{c,-}=0.94$.  These two filling factors represent the critical filling factor needed to stabilize the IQHWS: the IQHWS will only form at $\nu > \nu_{c,+}=1.06$ and $\nu < \nu_{c,-}=0.94$.
The $R_{xx}$ resistance peaks at these two filling factors, seen in Fig.1, are fairly wide in magnetic field, suggesting therefore a AI-IQHWS crossover rather than a sharp phase transition. These two critical filling factors were found to be nearly temperature-independent \cite{sean}. 

\begin{figure}[t]
  \includegraphics[width=\columnwidth]{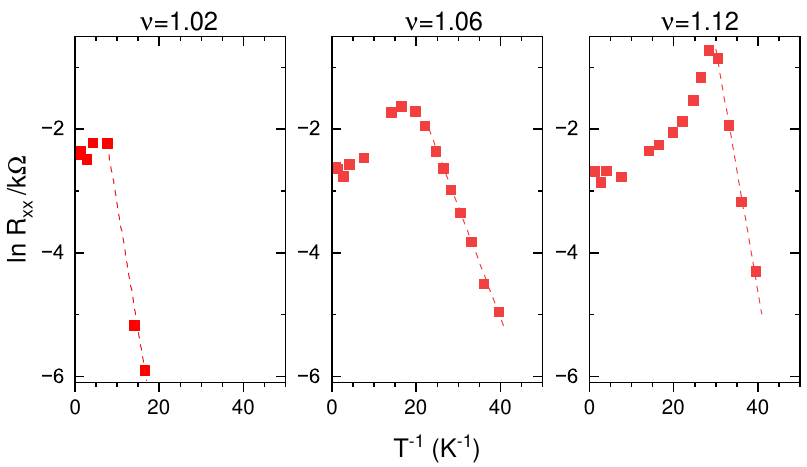} %5.2inch
  \caption{Arrhenius plots of $R_{xx}$ at selected values of the Landau level filling factor $\nu =1.02, 1.06,$ and $1.12$ measured in Sample 1. Dashed lines are fits to the low temperature parts of the data in the range of temperatures where $\ln R_{xx}$ exhibits a linear dependence on $1/T$. The slope of these dashed lines is a measure of the activation energy $E_a$.
\label{fig:2} }
\end{figure}

\begin{figure}[t]
  \includegraphics[width=\columnwidth]{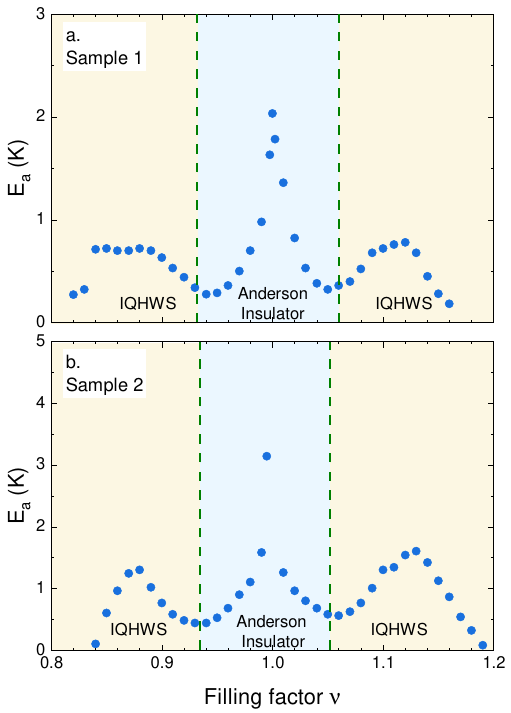} %5.2inch
  \caption{Activation energy $E_a$ in the vicinity of $\nu=1$ as plotted against $\nu$. Panels (a) and (b) represent data from Sample 1 and 2, respectively. Blue and yellow shaded regions correspond to the stability regions of the AI and IQHWS, respectively. The AI-IQHWS boundaries, as determined from transport measurements, are marked by the vertical dashed lines.
  The cusp of $E_a$ at $\nu=1$ is located in the center of the AI and the dome-like regions overlap well with the stability range of IQHWS. Furthermore, the two conspicuous local minima of the activation energy develop at the AI-IQHWS boundaries.
\label{fig:3}}
\end{figure}

When measured at sufficiently low temperatures, the totality of experimental results for the $\nu=1$ integer plateau
in samples of sufficiently high mobility \cite{chen,liu-1,liu-2,sean,muraki,smet,suslov,ashoori,liu-cap,young} and our data shown in Fig.1 is consistent with the sequence IQHWS-AI-IQHWS of phases. In contrast, in the limit of zero disorder and at finite temperatures, the IQHWS-uniform electron fluid-IQHWS sequence of phases is expected \cite{kivel}. The hallmark property of the uniform electron fluid forming in the middle of the plateau very close to integer Landau level filling factors is a classical Hall resistance \cite{kivel}. In this region our measurements do not exhibit such a classical Hall line, we thus do not observe the uniform electron fluid predicted for the zero disorder limit. Instead of a uniform electron fluid, close to $\nu=1$ disorder in our sample generates the AI, a phase characterized by a quantized Hall resistance. 

In the following we probe the low-lying thermal excitations both in the random and the collective localization regimes, i.e. for both the AI and the IQHWS. The linear parts of the $\ln R_{xx}$ versus $1/T$ curves shown in  Fig.\ref{fig:2} demonstrate that the longitudinal magnetoresistance is of the form $R_{xx} \sim \exp(-E_a/ k_\text{B} T)$. Here $E_a$ is the activation energy of the lowest lying thermal excitations. This means that transport in Sample 1 is
thermally activated across the whole $\nu=1$ plateau. For the AI, the thermally activated behavior is
well-known \cite{iqhe-ea1,iqhe-ea2}. Here we report that transport is also thermally activated for the IQHWS and also in the crossover region between the AI and the IQHWS. 

Next we focus on the dependence of the activation energy on the filling factor. 
Data shown in Fig.\ref{fig:2} suggest that this dependence is non-monotonic. Indeed, the slope of the linear region of the $\ln R_{xx}$ versus $1/T$ curve at $\nu=1.06$ is clearly smaller than that at either $\nu=1.02$ or $\nu=1.12$. The dependence of $E_a$ on $\nu$ for Sample 1, as measured along the $\nu=1$ plateau, is shown in Fig.\ref{fig:3}a. The activation energy curve is found to exhibit three local maxima. The largest activation energy is reached at $\nu=1$, in the center of the integer quantum Hall plateau. As the filling factor moves away from $\nu=1$, the activation energy plummets. Such a behavior is commonly observed in the integer quantum Hall regime and it reflects the density of states of localized quasiparticles \cite{iqhe-ea1,iqhe-ea2}. 
There are two other local minima of $E_a$ in the flanks of the integer plateau, at $\nu_{min,-}=0.94$ and $\nu_{min,+}=1.05$. Outside the $\nu_{min,-} < \nu < \nu_{min,+}$ range of filling factors, the activation energy exhibits an anomalous behavior. For example, in the $\nu_{min,+} < \nu < 1.16$ range, $E_a$ exhibits a dome-like shape, reaching a local maximum near $\nu \approx 1.12$. There is a similar behavior for hole-type quasiparticles in the $0.80 < \nu < 0.94$ range, where $E_a$ is also shaped like a dome. Therefore, the overall behavior of activation energy along the $\nu=1$ plateau can be described with the presence of a sharp cusp in the center of the plateau and two side lobes of local energy maxima in the flanks of the plateau. $E_a$ measured in Sample 2 is shown in Fig.\ref{fig:3}b and it exhibits a qualitatively similar behavior, exhibiting both the central cusp and the two side lobes. Quantitative differences of $E_a$ in these two samples will be discussed elsewhere.
We note that randomly localized quasiparticles do not exhibit the enhancements of $E_a$ in the flanks of integer quantum Hall plateaus \cite{iqhe-ea1,iqhe-ea2}. Furthermore, $E_a$ in Fig.\ref{fig:3} exhibits relatively large changes. In contrast, transport shown in Fig.\ref{fig:1} has relatively modest variations across the $\nu=1$ plateau.

A particularly interesting property of the activation energy is its correlation with the stability regions of the different phases. These correlations can be readily observed in Fig.\ref{fig:3} in both measured samples. Vertical lines in this figure mark the nearly temperature-independent boundaries $\nu_{c,\pm}$ between the AI and the IQHWS, as measured at $25$~mK. 
These boundaries align well with the minima of the activation energy occurring at $\nu_{min,\pm}$. We thus find that $\nu_{min,\pm} \approx \nu_{c,\pm}$, to within an error of about 1\%. 
The cusp at $\nu=1$ is observed in the central part of the AI. Similarly, the dome-shaped side lobes of $E_a$ overlap well with the stability range of the IQHWS, the regions shaded in yellow in Fig.\ref{fig:3}. We therefore associate the the dome-shaped side lobes of the activation energy with the IQHWS. According to one interpretation, the activation energy of the Wigner solid reflects its ability to move past pinning centers that are due to the disorder \cite{boris-1992,fertig-0}. We notice that the boundary between the AI and the IQHWS as well as locations of the minima of the activation energy are nearly sample-independent for the two samples we measured.

In the skyrmion interpretation, the sharp cusp in $E_a$ centered onto $\nu=1$ reflects the dependence of the activation energy of the randomly localized skyrmion excitations on the density of the skyrmions. Furthermore, we suggest that the enhanced $E_a$ lobes are due to a skyrmion crystal. The non-monotonic dependence of the activation energy on $\nu$ can therefore be interpreted as a crossover between the randomly localized skyrmions, or an Anderson insulator of skyrmions, and a skyrme crystal.

The development of deep local minima of the activation energy at filling factors $\nu_{min,\pm}$ at the boundary between the AI and the IQHWS remains our most surprising findings. This region could exhibit unusual phenomena, such as a structural phase transition in the skyrmion crystal \cite{cote-crystal-theory2,Rao1997} or the formation of a skyrmion glass \cite{disorder2}.
We conjecture that these minima are only present in samples of sufficiently low disorder, since the magnetoresistance features in the $R_{xx}(B)$ traces, such as the ones shown in Fig.1,  we associated with the boundary of the AI-IQHWS were not seen in work published prior to 2012 \cite{liu-1,liu-2,sean}. 
We think that these minima of the activation energy at $\nu_{min,\pm}$ cannot solely be accounted for by spin physics associated with skyrmions. 
The reasonably good coincidence of $\nu_{min,\pm}$ and $\nu_{c,\pm}$ suggests that the local minima in $E_a$ are localization driven to a large extent.
However, the understanding of these deep minima in the activation energy at the AI-IQHWS boundary and the behavior of the AI and IQHWS phases on the $\nu=1$ plateau remain interesting but unresolved problems closely connected to skyrmion physics. 

Recent state-of-the-art numerical work in the integer quantum Hall regime takes into account both disorder and many-body  effects \cite{osw1,osw2,osw3,osw4}. The exchange energy near $\nu=1$ is enhanced, but there appears to be no special signatures in this energy at $\nu=1.05$ and $0.94$, i.e. at the filling factors at which we observe deep energy minima associated with the boundary between the AI and the IQHWS \cite{osw1,osw2}.
For now, these calculations focused largely on charge ordered phases of high Landau levels \cite{osw3} and did not yet examine the region of small partial filling factors of about 0.05 very close to integer values nor skyrmionic effects near $\nu=1$.

Even though $\nu_{c,\pm}$ and the related $\nu_{min,\pm}$ are not understood, there is 
the following simplified, heuristic description for the crossover from random localization to charge order. As the quasiparticle density increases, quasiparticles will occupy the lowest energy bound states in the disorder potential landscape. This process has an overall smoothing effect of this potential as experienced by a further addition of new quasiparticles. As a result, the quasiparticles added past a critical density experience a significantly lower disorder, increasing therefore the likelihood of formation of charge order associated with the Wigner solid \cite{disorder2}. We note, however, that this simplified, heuristic description of the crossover from the AI to the IQHWS raises an interesting question on the dependence of the size of the skyrmions on their density. 
Hartree-Fock calculations in the limit of no disorder report a decreasing skyrmion size as the filling factor deviates from $\nu=1$ \cite{brey-crystal-theory1,cote-crystal-theory2}. In contrast, the simplified heuristic description presented above suggests that effects of the disorder are lower at large quasiparticle densities at which the skyrmion crystal forms. Lower disorder is associated with larger skyrmions \cite{disorder2, disorder1}. It appears, therefore, that there are two competing effects that determine the skyrmion size.

To conclude, we have conducted a study of the temperature dependence of the small signal transport within the $\nu=1$ quantum Hall plateau in samples from one of the latest generations of high-mobility 2DEGs. 
%Our findings exhibit transport patterns that can be associated with topological phases with different bulk insulators: the AI and the IQHWS. 
We found that the activation energy displays a very dramatic dependence on the magnetic field, a dependence that strongly correlates with the stability regions of the AI and the IQHWS of quasiparticles. In particular, the activation energy has local maxima in the central part of the AI and IQHWS stability regions and displays conspicuous local minima at the boundary between the AI and the IQHWS. We discussed the implications of our observations pertaining to skyrmion physics.

We acknowledge insightful discussions with Yi Huang and B.I. Shklovskii. Low temperature
measurements at Purdue University were supported by
the US Department of Energy award DE-SC0006671. 
The Princeton University portion of this research is funded in part by the Gordon and Betty Moore Foundation’s EPiQS Initiative, Grant GBMF9615.01 to Loren Pfeiffer.

\end{document}